\documentclass[journal]{IEEEtran}
\usepackage{graphicx}
\usepackage{hyperref}
\usepackage[tight,footnotesize]{subfigure}

\begin{document}
\title{LHC top mass: alternative methods and prospects for the future }

\author{Min Suk Kim for the ATLAS and CMS collaborations
\thanks{Min Suk Kim is with Kyungpook National University, Daegu, Korea.}
}

\maketitle
\pagestyle{plain}
\thispagestyle{empty}

\begin{abstract}
Future lepton colliders will have a rich top quark physics program which would add to our understanding of this interesting quark and promise an ultimate precision on the experimental knowledge of top-quark mass. We present possible improvements in the understanding of systematic uncertainties for various top-quark mass measurement techniques at the LHC, including projections for running conditions foreseen in the future.
\rm{\textit{Talk presented at the International Workshop on Future Linear Colliders (LCWS13), Tokyo, Japan, 11-15 November 2013.}}
\end{abstract}


\vspace{-0.3cm}
\section{Introduction}

In the short and mid-term future, top quark studies will be mainly driven by the Large Hadron Collider (LHC) experiments~\cite{CMS,ATLAS}. Exploration of top quarks will, however, be an integral part of particle physics studies at any future facility~\cite{Snowmass}. 
It is therefore an interesting question how much improvement in the measurement of the top-quark mass, $m_{t}$, can be accomplished at the LHC. We present possible improvements both for running conditions foreseen in the near and medium future and for various $m_t$ measurement techniques developed at the LHC.

Projections are provided for the expected integrated luminosity of 30 fb$^{-1}$, 300 fb$^{-1}$, and 3000 fb$^{-1}$ at a center-of-mass energy, $\sqrt{s}$ = 13-14 TeV. For the first two scenarios the pileup conditions are assumed to be somewhat elevated, while for the latter ({\it ``high-luminosity LHC or Phase II"}) scenario the pileup is assumed to be much more severe, requiring an upgraded detector to maintain good physics performance~\cite{FTR-13-017}.

With the dataset up to 5 fb$^{-1}$ at 7 TeV the LHC experiments have already achieved results comparable in precision to those from the Tevatron collider. The Tevatron experiments have reported a combined precision of 0.87 GeV (0.50\%)~\cite{Tevatron} and the LHC experiments have reached a precision of 0.95 GeV (0.55\%)~\cite{TOP-13-005}.
The dominant contributions to systematic uncertainty are the jet-energy scale (JES) calibration and QCD modeling.

Recently, the first world combination of $m_t$ measurements has been performed by the CDF and D$\O$ experiments at the Tevatron and the ATLAS and CMS experiments at the LHC. The resulting combined measurement is 173.34 $\pm$ 0.27 (stat.) $\pm$ 0.71 (syst.) GeV, with a total uncertainty of 0.76~GeV (0.44\%)~\cite{WorldComb}.
Since the experimental sensitivity is entering the sub-GeV range, issues of theoretical interpretation have become important.

Existing measurements rely on complex techniques required by the difficult hadron collider environment, while measurements at a lepton collider use either the invariant mass or a threshold scan that can give a clean theoretical interpretation of the result. New experimental techniques developed at the LHC help to mitigate this issue, but they typically have larger statistical uncertainties. Thus we estimate how the alternative methods may profit from large integrated luminosities. 

The extrapolations presented here are based on optimistic assumptions, anticipating improvements in the understanding of systematic uncertainties that are believed to be realistically possible with the large data samples. 
They also assume no new irreducible systematic uncertainties, as well as no known limit of experimental uncertainties. 
A lot of detailed study and dedicated work will be needed to achieve an understanding sufficient to claim the level of precision presented here~\cite{FTR-13-017}.

\begin{table*}[!t]
\caption{\label{tab:standard} Projection of the top-quark-mass precision (in GeV) obtained with standard methods based on reconstruction of the invariant mass of top-quark decay products, for various integrated luminosities using the assumptions explained in the text~\cite{FTR-13-017}.}
\noindent \centering{}%
\begin{tabular}{l|c|ccc|c}
 & Current
 & \multicolumn{3}{c|}{Future}
 & \multicolumn{1}{c}{Comment}\tabularnewline
Center-of-mass energy & 7 TeV & 13 TeV & 14 TeV & 14 TeV & \tabularnewline
                      & l+jets &   & & & \tabularnewline
Integrated luminosity & 5 fb$^{-1}$ & 30 fb$^{-1}$ & 300 fb$^{-1}$ & 3000 fb$^{-1}$ 
 &  
\tabularnewline
\hline
\hline

Fit calibration
 & 0.06 & 0.03 & 0.03 & 0.03 &  MC statistics \tabularnewline
b-JES
 & 0.61 & 0.27 & 0.09 & 0.03 & 3D fit \tabularnewline
Residual JES
 & 0.28 & 0.28 & \em 0.2 & \em 0.06 & {\em differential} \tabularnewline
Lepton energy scale
 & 0.02 & 0.02 & 0.02 & 0.02 & unchanged \tabularnewline
Missing transverse momentum
 & 0.06 & 0.06 & 0.06 & 0.06 & unchanged\tabularnewline
Jet energy resolution
 & 0.23 & 0.23 & \em 0.2 & \em 0.06 & {\em differential} \tabularnewline
b tagging
 & 0.12 & 0.06 & 0.06 & 0.06 & factor 2 (data) \tabularnewline
Pileup
 & 0.07 & 0.07 & 0.07 & 0.07 & unchanged \tabularnewline
Non-ttbar background
 & 0.13 & 0.06 & 0.06 & 0.06 & factor 2 (S/B) \tabularnewline
Parton distribution functions
 & 0.07 & 0.04 & 0.04 & 0.04 & factor 2 (PDF fits)\tabularnewline
Renormalization and factorization scales
 & 0.24 & 0.12 & 0.12 & \em 0.06 & full NLO + {\em differential} \tabularnewline
ME-PS matching threshold
 & 0.18 & 0.09 & 0.09 & \em 0.06 & full NLO + {\em differential} \tabularnewline
Underlying event
 & 0.15 & 0.15 & 0.15 & {\em 0.06} & {\em differential} \tabularnewline
Color reconnection effects
 & 0.54 & 0.27 & \em 0.2 & \em 0.06 & factor 2 + {\em differential} \tabularnewline
\hline
\hline
Systematic  & 0.98 & 0.60 & 0.44 & 0.20 &  \tabularnewline
Statistical & 0.43 & 0.15 & 0.05 & 0.01 &  \tabularnewline
\hline
Total & 1.07 & 0.62 & 0.44 & 0.20 & \tabularnewline
\end{tabular}
\end{table*}

\section{Conventional methods}

ATLAS and CMS demonstrated improved methods to reduce the systematic uncertainty by using the 3-dimensional analysis~\cite{3dTemplate} and differential analysis~\cite{TOP-12-029}, respectively. In addition, updated measurements based on the 2012 LHC run at 8 TeV are being performed. Future measurements and combinations can lead to a reduction of systematic uncertainties with improved analysis techniques and high-luminosity LHC, even in the high pileup environment. 

Conventional measurements of $m_{t}$ at hadron colliders are based on the measurement of the invariant mass of the top-quark decay products. The recent measurements by CMS in the $\rm{t}\bar{\rm{t}} \rightarrow$ lepton+jets, $\rm{t}\bar{\rm{t}} \rightarrow$ di-lepton and $\rm{t}\bar{\rm{t}} \rightarrow$ all jets channels have been done by this type of analysis.
The possible evolution in the systematic uncertainties using conventional technique is first investigated. The extremely high pileup expected for the 3000 fb$^{-1}$ data makes predictions very uncertain. 

In this analysis note~\cite{FTR-13-017}, we summarized core assumptions for the extrapolation of the conventional method and highlighted the most promising methods to reduce the systematic uncertainties in the future. The $\rm{t}\bar{\rm{t}}$ cross-section increases by a factor 5.7 between 7 and 14 TeV, and this increased $\sigma_{t\bar{t}}$ will compensate for possible losses in sensitivity due to trigger conditions and increased pileup, which means that uncertainties will just scale with increased luminosity, compared to 7 TeV.
The published result from the CMS collaboration in the $\rm{t}\bar{\rm{t}} \rightarrow$ lepton+jets channel using 5 fb$^{-1}$ of data at 7 TeV~\cite{JHEP12} is used as a baseline for the projections of the expected precision at various 13-14 TeV scenarios.

The breakdown of the systematic uncertainty in its main components is shown in Table~\ref{tab:standard}. The dominant contributions are related to the JES calibration and resolution, non-perturbative QCD effects like color reconnection and underlying event, as well as perturbative QCD parameters, in particular the renormalization and factorization scale ($Q^2$) and the threshold used for the matching between matrix-element (ME) level and parton showering (PS).

For the 300~fb$^{-1}$ scenario we assume that we will fully profit from next-to-leading-order (NLO) simulation and detailed data-based constraints on hard and soft QCD, as well as dedicated JES calibrations and 3D fit. The 3D analysis uses a 3-dimensional template technique which determines $m_{t}$ together with a global jet energy scale factor, and a relative b-jet to light-jet energy scale factor (b-JES), where light jets refer to u, d, c, s quark jets.
At 3000 fb$^{-1}$ we can fully benefit from the 3D and differential analysis approaches to further limit the uncertainties.

\section{Alternative methods}

Alternative measurement methods based on observables that can be calculated in theoretically well-defined mass schemes can ultimately lead to a better precision, though even such approaches would eventually be limited by $\Lambda_{\rm{QCD}}$~\cite{FTR-13-017}.

\subsection{The endpoint method}

The end-point method~\cite{endpoint} is based on the observation that the invariant mass distribution of a lepton and a b-jet contains a relatively sharp edge whose position is correlated with $m_{t}$. The attraction of measuring the position of this end-point is that it is (almost) independent of any assumption about the matrix element. That is to say it uses the end-point of distributions, rather than the mean value, without the use of Monte Carlo (MC) simulation to calibrate the result.

This analysis, however, is sensitive to the b-JES in a very similar way as the standard measurements. As we explain below, other alternative approaches such as the $L_{xy}$ and $J/\psi$ method are not sensitive to the b-JES. Note, however, that they rely on a good understanding of the B-hadron fragmentation functions~\cite{TOP-13-007}.

For the projections shown in Table~\ref{tab:endpoint}, we optimistically assume that the uncertainties related to the background
modeling and fit range can be reduced with more data, scaling down as a statistical uncertainty.
Since the JES cannot be fitted in-situ with this analysis we assume that the improved knowledge from other measurements can be useful for this analysis to a certain extent. As this will inevitably introduce an extrapolation uncertainty for possible differences in phase space, jet flavor and the fact that the endpoint method relies on the absolute JES in data, rather than the relative difference between data and MC, we quote 0.3 GeV as ultimate residual uncertainty for 3000 fb$^{-1}$~\cite{FTR-13-017}.

\begin{table*}
\caption{\label{tab:endpoint} Projection of the top-quark mass precision (in GeV) obtained with the endpoint method, for various integrated luminosities using the assumptions explained in the text~\cite{FTR-13-017}.}
\noindent \centering{}%
\small
\begin{tabular}{l|c|ccc|c}
 & Current
 & \multicolumn{3}{c|}{Future}
  & \multicolumn{1}{c}{Comment}\tabularnewline
Center-of-mass energy  & 7 TeV & 13 TeV & 14 TeV & 14 TeV & \tabularnewline
Integrated luminosity & 5 fb$^{-1}$ & 30 fb$^{-1}$ & 300 fb$^{-1}$ & 3000 fb$^{-1}$ & \tabularnewline
\hline
\hline
Jet energy scale and resolution
 & 1.6 & 0.9 & 0.5 & 0.3 & improve with data \tabularnewline
Lepton energy scale
 & 0.4 & 0.2 & 0.2 & 0.2 & factor 2 \tabularnewline
Jet and lepton efficiencies
 & 0.2 & 0.2 & 0.2 & 0.2 & unchanged \tabularnewline
Fit range
 & 0.6 & 0.2 & 0.2 & 0.2 & statistics (factor 4) \tabularnewline
Background shape
 & 0.5 & 0.2 & 0.1 & 0.02 & statistics \tabularnewline
QCD effects
 & 0.6 & 0.3 & 0.3 & 0.3 & factor 2 \tabularnewline
Pileup
 & 0.1 & 0.1 & 0.1 & 0.1 & unchanged \tabularnewline
\hline
\hline
Systematic  & 1.9 & 1.0 & 0.6 & 0.5 &  \tabularnewline
Statistical & 0.9 & 0.4 & 0.1 & 0.04 & \tabularnewline
\hline
Total       & 2.1 & 1.1 & 0.6 & 0.5 & \tabularnewline
\end{tabular}
\end{table*}

\subsection{The $J/\psi$ method}

The $J/\psi$ method uses the invariant mass of a $J/\psi$ from the b-jet together with the lepton from the $W$ decay as observable. 

Preliminary studies with $t \rightarrow Wb \rightarrow l\nu + (J/\psi + X)$ candidates from the data collected at 8 TeV~\cite{TOP-13-007} already strongly favor the event rate at 14 TeV, based on $J/\psi \rightarrow \mu^{+}\mu^{-}$ decay alone.
The observed rate at 8 TeV as well as the main kinematic distributions of these rare decays are in reasonable agreement with the current default simulation.

Table~\ref{tab:Jpsi} shows the projections as a function of luminosity, with the uncertainties adopted from the CMS Physics TDR study~\cite{TDR,Jpsi} as a starting point. We assume, as in the previous sections, that the systematic uncertainties can be further reduced over time, partly using constraints from the data itself, partly based on theory developments. The dominant remaining systematic uncertainties will be related to the QCD scales and the b-fragmentation functions.

\begin{table*}
\caption{\label{tab:Jpsi} Expected top-quark mass precision (in GeV) achieved with the $J/\psi$ method, for various integrated luminosities using the assumptions explained in the text~\cite{FTR-13-017}.}
\noindent \centering{}%
\small
\begin{tabular}{l|ccc|c}
 & \multicolumn{3}{c|}{Future}
 & \multicolumn{1}{c}{Comment}\tabularnewline
Center-of-mass energy & 13 TeV   & 14 TeV    & 14 TeV     & \tabularnewline
Integrated luminosity & 30 fb$^{-1}$ & 300 fb$^{-1}$ & 3000 fb$^{-1}$ & \tabularnewline
\hline
\hline
Parton distribution functions
 & 0.3 & 0.2 & 0.1 & improve with theory and data \tabularnewline
Renormalisation and factorization scales
 & 0.9 & 0.4 & 0.4 & improve with NNLO for $m_{\ell B}$ \tabularnewline
Initial- and final-state radiation
 & 0.3 & 0.2 & 0.1 & full NLO gen. + diff. data \tabularnewline
b and light fragmentation
 & 0.7 & 0.5 & 0.3 & improve with data \tabularnewline
Underlying event
 & 0.6 & 0.2 & 0.1 & improve with data \tabularnewline
Lepton energy scale and resolution
 & 0.5 & 0.2 & 0.2 & improve with data \tabularnewline
Jet energy scale and resolution
 & 0.1 & 0.1 & 0.1 & \tabularnewline
Background knowledge
 & 0.2 & 0.1 & 0.1 & \tabularnewline
\hline
\hline
Systematic  & 1.5 & 0.8 & 0.6 & \tabularnewline
Statistical & 1.0 & 0.3 & 0.1 & \tabularnewline
\hline
Total       & 1.8 & 0.8 & 0.6 & \tabularnewline
\end{tabular}
\end{table*}

\begin{table*}
\caption{\label{tab:Lxy} Projection of the top-quark mass precision (in GeV) obtained with the $L_{xy}$ method, for various integrated luminosities using the assumptions explained in the text~\cite{FTR-13-017}.}
\noindent \centering{}%
\small
\begin{tabular}{l|c|ccc|c}
 & Current
 & \multicolumn{3}{c|}{Future}
 & \multicolumn{1}{c}{Comment}\tabularnewline
Center-of-mass energy & 8 TeV & 13 TeV & 14 TeV & 14 TeV & \tabularnewline
                      & $e\mu$  &        &        &        & \tabularnewline
Integrated luminosity & 20 fb$^{-1}$ & 30 fb$^{-1}$ & 300 fb$^{-1}$ & 3000 fb$^{-1}$ & \tabularnewline
\hline
\hline

b fragmentation/hadronization
 & 0.8 & 0.4 & 0.4 & 0.3 & improve with data\tabularnewline
Top $p_T$ modeling
 & 2.4 & 0.2 & 0.2 & 0.2 & improve at NNLO \tabularnewline
Other systematic uncertainties
 & 1.1 & 0.3 & 0.2 & 0.2 & improve with data\tabularnewline
\hline
\hline
Systematic  & 2.8 & 0.6 & 0.5 & 0.4 &  \tabularnewline
Statistical & 2.0  & 1.1 & 0.4  & 0.1 & \tabularnewline
\hline
Total & 3.4 & 1.3 & 0.6 & 0.4 & \tabularnewline
\end{tabular}
\end{table*}

\subsection{The $L_{xy}$ method}

The $L_{xy}$ method~\cite{Lxy} relies on the decay length of B hadrons from $\rm{t}\bar{\rm{t}}$ decays and avoids the reconstruction of jets, reducing strongly the traditional uncertainties related to the JES calibration and QCD processes and possibly reducing dependence on pileup.

The CDF collaboration pioneered this method~\cite{Lxy-CDF}, and CMS recently used and reached sub-percent relative statistical precision for the first time, combining lepton+jets and di-lepton channels~\cite{TOP-12-030}.

The dominant sources of systematic uncertainties are the b-fragmentation functions, like the $J/\psi$ method, and the modeling of the top-quark $p_T$ which is not well understood since it is based on the b-quark boost~\cite{Differential}. 
We optimistically assume that this effect can be substantially reduced from a full NNLO computation of this quantity as reported in~\cite{Computations}.
We also assume substantial improvement for the remaining systematic uncertainties mostly related to the signal modeling.
Using the $e\mu$ channel, we simplified the projections as shown in Table~\ref{tab:Lxy}.

\subsection{Extraction from the cross section}

Results with standard or alternative methods can be cross-checked by determining the top mass from a well-defined theoretical mass scheme. Comparison of the measured $\rm{t}\bar{\rm{t}}$ cross section to the theoretical prediction has the advantage and already been used at the Tevatron and LHC. The D$\O$ collaboration first used this method~\cite{Xsec-D0} and ATLAS at 7 TeV in the early stage~\cite{ATLAS-CONF-2011-054}.

Using the latest $\rm{t}\bar{\rm{t}}$ cross section predictions at NNLO + NNLL QCD~\cite{Czakon}, CMS recently obtained the pole mass with an uncertainty of slightly more than 3 GeV (2\%), under the assumption that $m_t^{MC}$ = $m_t^{pole} \pm 1.00$ GeV at fixed value of $\alpha(m_Z)$ = 0.1184 $\pm$ 0.0007 which is the latest world average value with its corresponding uncertainty~\cite{Xsec-CMS}.

It should be noted that this method is expected to be limited by relatively poor sensitivity of $\sigma_{t\bar{t}}$ to $m_t$. With the optimistic assumption that the acceptance does not depend on the $m_t$, the uncertainty can be reduced to just below 2 GeV.

\subsection{Future alternative}

As a future alternative, we also discuss the $m_t$ measurement using Bi-Event Subtraction Technique (BEST) proposed~\cite{BEST}.
This method is for modeling and subtracting dominant combinatorial background during reconstruction of particle decay chains at hadron colliders. The background shape can be modeled by combining observed particles from different collision events.

The basic idea of BEST can be demonstrated for the reconstruction of the $W$ boson decaying into two jets at the LHC. In a sample of jet pairs collected for each event, some of jet pairs may come from a single $W$ boson decay in the events, while other jet pairs will be combinatorial background. By collecting another sample of jet pairs where each jet comes from a different event we can model this combinatorial background well.
A general application can reconstruct the top quark while combining the b-jet with the reconstructed $W$ in $t \rightarrow Wb \rightarrow jjb$.

From preliminary studies, we conclude that a promising approach to this method is based on the reconstructed mass ratio, $R = m_{jjb}/m_{jj}$ to minimize the impact of the JES uncertainty on $m_t$. Here $m_{jjb}$ and $m_{jj}$ are the per-event invariant masses of the hadronically decaying top quark and $W$ boson reconstructed from three and two jets respectively.

The analysis, based on the observable $R$, performed by ATLAS utilises a kinematic fit maximising an event likelihood to select the jet triplet for determining the two masses~\cite{R32}, while the analysis being performed by CMS doesn't, but both analyses use the template method in the $\rm{t}\bar{\rm{t}} \rightarrow$ lepton+jets channel.

\section{Conclusions}

\begin{figure}[!t]
  \centering
    \includegraphics[width=.5\textwidth]{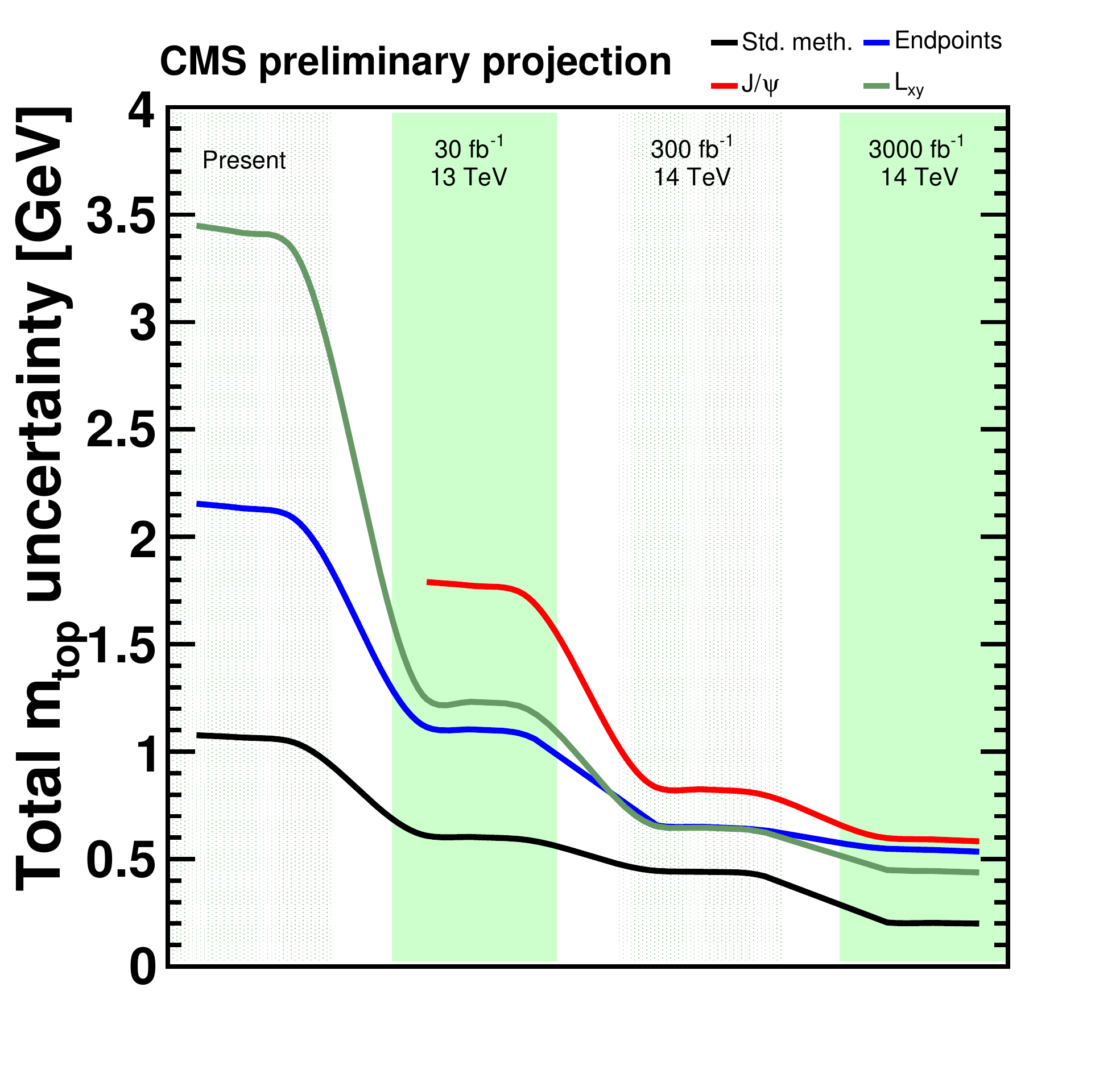}
    \caption{Projection of the top-quark-mass precision obtained with different measurement methods, for various integrated luminosities.}
    \label{fig:Fig1}
\end{figure}

We have evaluated the potential ultimate precision on the top-quark mass both for standard methods and for several promising and useful alternative approaches. Please note that the assumptions used in the extrapolations to high luminosity at various 13-14 TeV scenarios are cautiously optimistic. A summary is shown in Table~\ref{tab:Summary} and Fig.~\ref{fig:Fig1}.

\begin{table}[!b]
\caption{\label{tab:Summary} Projection of the top-quark mass precision (in GeV) obtained with current methods, for various integrated luminosities using the assumptions explained in the text~\cite{FTR-13-017}.}
\noindent \centering{}%
\small
\begin{tabular}{l|c|ccc}
 & Current
 & \multicolumn{3}{c}{Future}\tabularnewline
$\sqrt{s}$  & 7 TeV & 13 TeV & 14 TeV & 14 TeV \tabularnewline
$\mathcal{L}_{integrated}$ & 5 fb$^{-1}$ & 30 fb$^{-1}$ & 300 fb$^{-1}$ & 3000 fb$^{-1}$ \tabularnewline
\hline
\hline
$J/\psi$ method
 & - & 1.8 & 0.8 & 0.6 \tabularnewline
$L_{xy}$ (8 TeV)
 & 3.4 & 1.3 & 0.6 & 0.4 \tabularnewline
Endpoints
 & 2.1 & 1.1 & 0.6 & 0.5 \tabularnewline
Standard method
 & 1.1 & 0.6 & 0.4 & 0.2 \tabularnewline
\hline
\hline
\end{tabular}
\end{table}

\vfill
~


%

\end{document}